\documentclass[10pt]{article}

\usepackage[top=3.0cm, bottom=3.0cm, outer=3.0cm, inner=3.0cm]{geometry}
\usepackage{graphicx}

\usepackage{amsfonts}
\usepackage{mathpazo}
\usepackage{amsmath}
\usepackage{physics}
\usepackage{mathtools}
\usepackage{bm}
\usepackage{setspace}
\usepackage{indentfirst}

\usepackage[dvipsnames]{xcolor}
\usepackage{hyperref}
\usepackage{cleveref}
\hypersetup{colorlinks={true}, allcolors={RoyalBlue}}
\usepackage[biblabel]{cite}
\usepackage{ftnxtra}

\usepackage{tikz}
\usetikzlibrary{calc} 
\usepackage[f]{esvect} 

\DeclareRobustCommand{\d}{\kern-.05em \reflectbox{\ensuremath{\vv{\reflectbox{$d$}}}} \kern-.15em}
\DeclareRobustCommand{\lvec}[1]{\reflectbox{\ensuremath{\vv{\reflectbox{$#1$}}}} \kern-.10em}
\DeclareRobustCommand{\vec}[1]{\kern-.10em \vv{#1} \kern-.10em}
\DeclareRobustCommand{\longvec}{\vv}
\newcommand{\pfrac}[3]{{{\left. \frac{\partial #1}{\partial #2} \right|}_{#3}}}
\newcommand{\pslash}[3]{{{\left. {\partial #1}/{\partial #2} \right|}_{#3}}}

\DeclareMathOperator{\Aut}{Aut}

\makeatletter
\def\@maketitle{%
    {\fontsize{17}{20}\selectfont \textbf{\@title} \par}
    \vskip 1.5em
    \list{}{
        \setlength{\leftmargin}{1.0cm}
        \setlength{\rightmargin}{0cm}
    }\item\relax
    {\@author \par}
    \endlist
    \vskip 1.0em}
\makeatother

\title{Thermodynamic identities with sunray diagrams}
\author{\large\textbf{Joon-Hwi Kim\textsuperscript{1} and Juno Nam\textsuperscript{2}}\\[1em]
\small{\textsuperscript{1} Department of Physics and Astronomy, Seoul National University, Seoul, South Korea\\
\textsuperscript{2} Department of Chemistry, Seoul National University, Seoul, South Korea\\[1em]
E-mail: joonhwi.kim@snu.ac.kr}}
\date{}

\renewenvironment{abstract}{
\list{}{
    \setlength{\leftmargin}{1.0cm}
    \setlength{\rightmargin}{0cm}
}
\item\relax
}{\endlist}

\begin{document}
\maketitle
\begin{abstract}
\textbf{Abstract}

One of the hurdles in teaching undergraduate thermodynamics is a plethora of complicated partial derivative identities.
Students suffer from difficulties in deriving, justifying, or interpreting the identities, misconceptions about partial derivatives, and a lack of in-depth understanding of the meanings of identities.
We propose a novel diagrammatic method for the calculus of differentials and partial derivatives called the ``sunray diagram'' that resolves the difficulties above.
The sunray diagram technique relates a partial derivative with ``arrow sliding,'' which enables an aesthetic and intuitive manipulation of partial derivative expressions in the form of successive arrow slidings.
Furthermore, the sunray diagram is more than an ad hoc or abstract machinery but is based on the symplectic structure of thermodynamics; the sunray diagram admits a direct physical interpretation on the $P$-$V$ (or $T$-$S$) plane.
We elaborate on such physical semantics of the sunray diagram by taking Maxwell's approach to the geometry of thermodynamic structures---reinterpreted in terms of differential geometry---as a reference point.
We anticipate that our discussion introduces the geometry of thermodynamics to learners and enriches the graphical pedagogy in physics education.

Keywords: thermodynamics, graphical notation, partial derivatives, Maxwell relations, symplectic geometry
\end{abstract}
\vspace{0.75cm}

\setlength{\parskip}{5pt}

\section{\label{sec:Intro}Introduction}

Becoming adept at partial derivative identities for deriving various thermodynamic identities is a vital mission to be completed in undergraduate thermodynamics.
However, some identities may seem unfamiliar to students or are heavy to be memorized, while it is difficult to justify or derive them without involving technical details about partial derivatives.
Students who are busy to catch up with the mathematical details may be missing the physical context of the equations; students who are well-acquainted with the mathematical aspects may also have weaknesses such as misconceptions about the physicists' manner of handling partial derivatives (which differs from that of standard mathematics texts) or exploiting the identities merely as formal manipulation rules but lacking deep understandings of their meanings.
Therefore, it will be pedagogically valuable to develop a tool that can intuitively derive the partial derivative identities, serve as a quick mnemonic for them, and deepen understandings and clarify concepts of the partial derivative system.
Then, both the practical users of the identities and those who lay emphasis on pursuing deeper conceptual understandings will benefit largely.
Also, it will be nicer if it is supported by a certain standard of mathematical rigor.
The ``sunray diagram'' presented in this paper successfully fulfills all these conditions as a graphical language for differentials and partial derivatives.

Attempts to utilize graphics for the calculus of thermodynamics can be traced back to J. C. Maxwell.
Maxwell, when developing his ``Theory of Heat,''\cite{maxwell1891theory} often visualized the equations to get physical and geometrical insights.
He graphically interpreted the partial derivatives as infinitesimal segments over contour lines of thermodynamic variables and did the calculus of differentials by applying successive equal-area sliding of an infinitesimal parallelogram.
Refer to Nash's article\cite{nash1964carnot} for a reproduction.
There are also other notable works that considered the geometrical interpretation or formulation of thermodynamics.
Gibbs's seminal works\cite{gibbs1948collected} described properties related to phase equilibrium, and some following works\cite{nyburg1961geometric,beste1961geometric,hantsaridou2006geometry} interpreted thermodynamic equations geometrically on the diagram of thermodynamic variables.
Also, there are contact-geometric descriptions\cite{hermann1973geometry,mrugala1978geometrical,bravetti2015contact} and metric-based approaches.\cite{ruppeiner1979thermodynamics,weinhold1975metric}
These works are insightful and capture the concept of partial derivatives directly and accurately.
However, they are not favorable for a graphical language in thermodynamics since they place importance on geometric interpretation rather than utilization or have complicated semantics of their geometric elements.
Weinhold\cite{weinhold1975metric,weinhold1976thermodynamics} sought a vector description of thermodynamic variables with arrows based on the metric structure, but the physical meanings of lengths and angles in the diagram are hard to be interpreted directly.

Meanwhile, the Jacobian technique\cite{crawford1949jacobian,carroll1965use,landau1980sp} interprets partial derivatives as a specific form of Jacobian and provides simple algebraic derivations for most of thermodynamic partial derivative identities.
Although a geometric picture can be supplemented with it by interpreting Jacobians as the scaling of an infinitesimal area under a coordinate transformation, the method is primarily based on algebraic manipulations rather than graphical reasoning.
Hence, it appears that the Jacobian technique is commonly used as a purely algebraic method.
On the other hand, the geometric approaches from Maxwell to Weinhold seem to be unsuccessful in being developed into comparably competitive graphical notations and tools.

The method developed in the paper---the ``sunray diagram''---alleviates the complexities of previous graphical methods for thermodynamics and provides a complementary geometric intuition to all, including the Jacobian technique.
The sunray diagram method involves a chain of sliding as Maxwell's method did, but this time, to be slid are arrows, not parallelograms, and deciding which sliding pathway should be taken to obtain the wanted identity is more straightforward.
In addition, graphical elements of a sunray diagram can be easily translated to ordinary mathematical expressions.
Also, as a sunray diagram can be endowed with the oriented area structure, the sunray diagram method incorporates the Jacobian technique in the graphical form.
Furthermore, the arrow sliding mechanism in a sunray diagram makes the method more convenient than the Jacobian technique in some cases in which complicated dependencies exist between variables.
In sum, the sunray diagram is a method primarily based on the vector intuition rather than the area intuition of previous approaches and makes some unforeseen shortcuts accessible.

Launching a new tool, we would like to provide a user's manual.
This paper is organized as follows.
Section \ref{sec:Sunray} explains the basic elements of the sunray diagram and demonstrates how to use it for quickly deriving the partial derivative identities in a layman-friendly manner.
The mnemonic aspects of the sunray diagram are presented with applications to some well-known partial derivative identities in undergraduate thermodynamics.
Then, in section \ref{sec:Exp}, the geometrical basis of the sunray diagram is concretely elucidated in the language of symplectic geometry, which is the underlying geometrical feature of thermodynamics.\cite{hermann1973geometry,mrugala1978geometrical,kocik1986geometry,schutz1980geometrical}
We take the aforementioned Maxwell's approach and investigate its relationship with the sunray diagram, and it turns out that the two are different graphical representations of the same mathematical structure.
From these discussions, it will be apparent that the sunray diagram is not an ad hoc machinery but has geometrical or physical interpretations so that it is practical and insightful at the same time.

\section{\label{sec:Sunray}The Sunray Diagram}
\subsection{\label{sec:SunrayBasic}Basic Syntax}

\begin{figure}[t]
    \centering
    \includegraphics{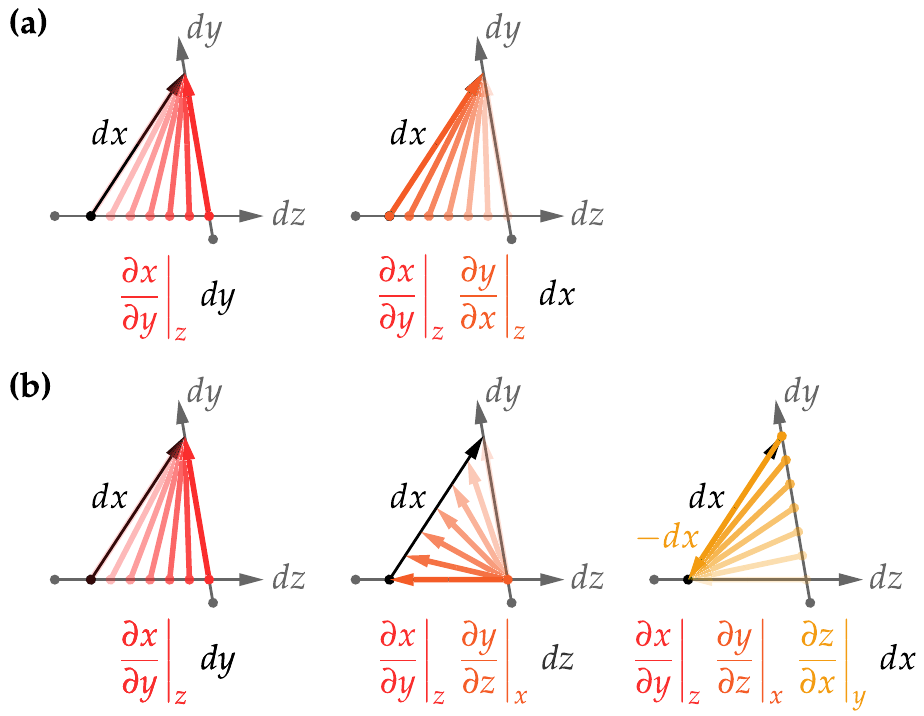}
    \caption{Sunray diagrams for proving equations \eqref{eq:reci} and \eqref{eq:triple}, respectively.}
    \label{fig:proof1}
\end{figure}

Suppose there are three variables $x$, $y$, and $z$ that each of them depends on the remaining two.
The partial derivatives appear as the coefficients of differentials:
\begin{equation}
    dx = \pfrac{x}{y}{z} \, dy + \pfrac{x}{z}{y} \, dz.
\label{eq:pdcoeff}
\end{equation}
The notation $\pslash{x}{y}{z}$ denotes the partial derivative of $x$ with respect to $y$ with the value of $z$ held constant.
The partial derivative is generally denoted by $(\partial x/\partial y)_z$ in thermodynamics literature (e.g., \cite{reif2009fundamentals,schroeder1999introduction,callen1998thermodynamics}), but we follow the notation used by Kardar \cite{kardar2007sp} for visual brevity.
Note that $\pslash{x}{y}{z} = dx/dy$ when $dz$ is set to zero in equation \eqref{eq:pdcoeff}.
It is well-known that the following identities hold:
\begin{align}
    \label{eq:reci}
    \pfrac{x}{y}{z} \, \pfrac{y}{x}{z} &= 1;\\
    \label{eq:triple}
    \pfrac{x}{y}{z} \, \pfrac{y}{z}{x} \, \pfrac{z}{x}{y} &= -1.
\end{align}
These identities are frequently used in thermodynamics.
For example, one can express $\pslash{P}{T}{V}$ in terms of thermal properties:
\begin{align}
    \label{eq:thermprop}
    \pfrac{P}{T}{V} =
    -\frac{1}{\pslash{T}{V}{P}\,\pslash{V}{P}{T}} =
    -\frac{\pslash{V}{T}{P}}{\pslash{V}{P}{T}} = -\frac{\alpha_P}{\kappa_T},
\end{align}
where $\alpha_P$ and $\kappa_T$ are the isobaric thermal expansion coefficient and the isothermal compressibility.
Equations \eqref{eq:reci} and \eqref{eq:triple} can be proved in the ordinary notation from
\begin{align}
    \label{eq:proof}
    dx &= \pfrac{x}{y}{z} \, dy + \pfrac{x}{z}{y} \, dz \nonumber \\
    &= \pfrac{x}{y}{z} \left( \pfrac{y}{x}{z} \, dx + \pfrac{y}{z}{x} \, dz \right) + \pfrac{x}{z}{y} \, dz,
\end{align}
by comparing the coefficients of $dx$ and $dz$ on both sides. However, the peculiar minus sign in equation \eqref{eq:triple} still remains mysterious: it is derived mathematically, but we do not have a mental picture.
Now, have a look at figure \ref{fig:proof1}: a visual justification of equations \eqref{eq:reci} and \eqref{eq:triple} is immediately obtained.

\begin{figure}[t]
    \centering
    \includegraphics{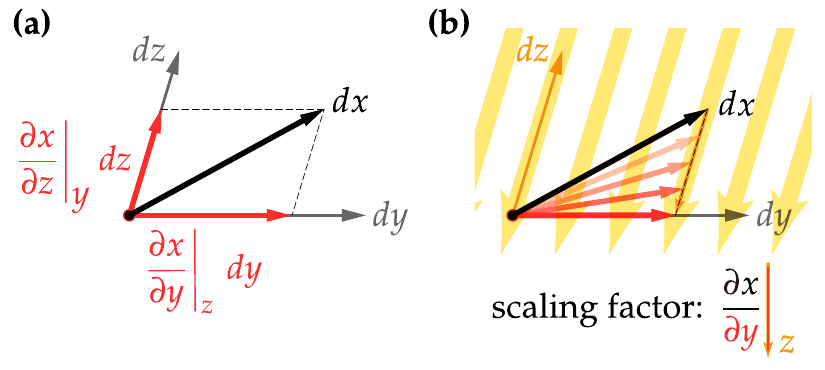}
    \caption{(a) The basic interpretation of a sunray diagram. (b) An alternative way to read a sunray diagram.}
    \label{fig:sunray0}
\end{figure}

\begin{figure}[t]
    \centering
    \includegraphics{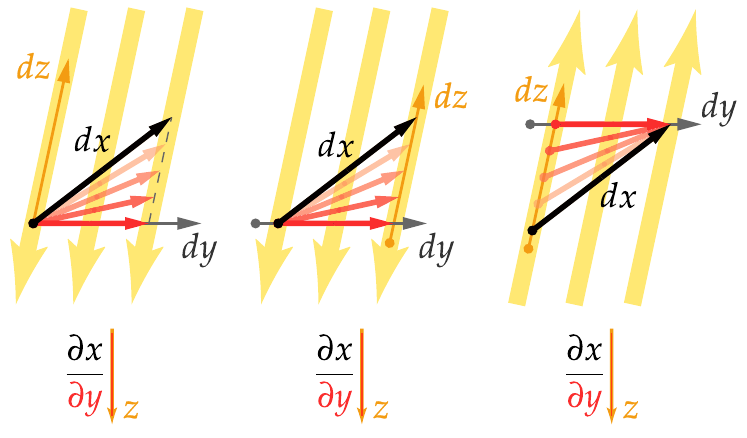}
    \caption{Sunray diagrams with different arrangements of vectors.}
    \label{fig:parallel}
\end{figure}

Figure \ref{fig:sunray0} explains how these diagrams work. 
Equation \eqref{eq:pdcoeff} can be interpreted as a decomposition of a ``vector'' $dx$ into components parallel to the ``vectors'' $dy$ and $dz$ (figure \ref{fig:sunray0}(a)), where the two ``vectors'' $dy$ and $dz$ are need not be orthogonal.
Accepting such an idea of graphically representing differentials as vectors, one can geometrically interpret $\pslash{x}{y}{z}$.
The ``vector'' $\pslash{x}{y}{z} \, {dy}$ is the shadow of the ``vector'' ${dx}$ when projected to the direction of ${dy}$ by a sunray parallel to ${dz}$, and the scaling factor of this projection is $\pslash{x}{y}{z}$ (figure \ref{fig:sunray0}(b)).
It is easy to remember such an assignment of a scaling factor to a movement in a diagram.
Sliding ${dx}$ to ${dy}$ along a $z$-sunray (a line parallel to the vector ${dz}$) translates into writing down $\partial x$ as a numerator and $\partial y$ as a denominator, then drawing a vertical line from $\partial x$ to $\partial y$ with a small ``$z$'' placed next to it. 
Lastly, note that such reading of an ``arrow sliding'' is regardless of how ``vectors'' in a diagram are placed, as illustrated in figure \ref{fig:parallel}. 
Since vectors can be arbitrarily moved by parallel translations, one can choose a convenient configuration when drawing a sunray diagram.


Now, return to equations \eqref{eq:reci} and \eqref{eq:triple}. 
The corresponding sunray diagram consists of a triangle formed by ${dx}$, ${dy}$, and ${dz}$, as drawn in figure \ref{fig:proof1}. 
In the first step of figure \ref{fig:proof1}(a), ${dx}$ is projected along a $z$-sunray to the $dy$-axis (a line parallel to ${dy}$).
The resulting arrow is $\pslash{x}{y}{z} \, {dy}$, the ${dy}$ term when ${dx}$ is written in terms of ${dy}$ and ${dz}$ (decomposed into directions parallel to arrows representing ${dy}$ and ${dz}$). 
In the next step, $\pslash{x}{y}{z} \, {dy}$ is again moved along a $z$-sunray and returns to ${dx}$ to be $\pslash{x}{y}{z} (\pslash{y}{x}{z} \, {dx})$.
Equating this with ${dx}$ proves equation \eqref{eq:reci}.
In figure \ref{fig:proof1}(b), $dx$ goes on an excursion: visiting the $dy$-axis, $dz$-axis, and then returning home. 
The net scaling factor it gains is $\pslash{x}{y}{z} \pslash{y}{z}{x} \pslash{z}{x}{y}$. 
This must be equal to $-1$ since its direction gets flipped after running a lap. 
This proves equation \eqref{eq:triple} and provides a visual intuition to the peculiar minus sign. 

\begin{figure}[b!]
    \centering
    \includegraphics{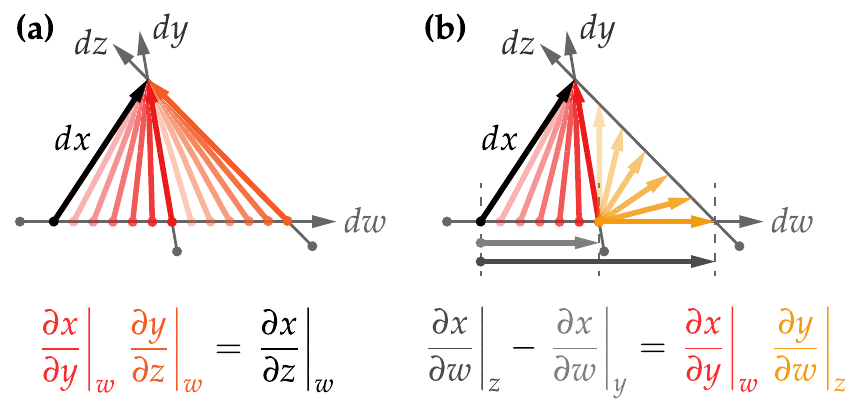}
    \caption{Sunray diagrams for proving equations \eqref{eq:chain1} and \eqref{eq:chain2}, respectively.}
    \label{fig:proof2}
\end{figure}

\begin{figure}[t!]
    \centering
    \includegraphics{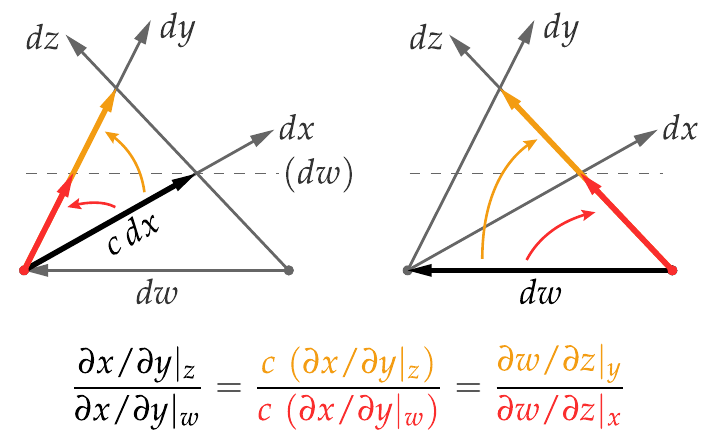}
    \caption{A sunray diagram for proving equation \eqref{eq:chain3}, where $c$ is a constant. Note that ``arrow slidings'' are indicated by curved arrows for the sake of visual brevity.}
    \label{fig:proof3}
\end{figure}

When a new member, $dw$, participates in this linear system of differentials, more partial derivative identities appear.
For example, the identities
\begin{align}
    \label{eq:chain1}
    \pfrac{x}{y}{w} \, \pfrac{y}{z}{w} &= \pfrac{x}{z}{w}, \\
    \label{eq:chain2}
    \pfrac{x}{w}{z} - \pfrac{x}{w}{y} &= \pfrac{x}{y}{w} \, \pfrac{y}{w}{z}
\end{align}
are used when deriving thermodynamic identities such as
\begin{equation}
    \label{eq:cp-cv}
    C_P - C_V = \frac{\alpha_P^2 TV}{\kappa_T},
\end{equation}
the identity about the difference between the isobaric and isochoric heat capacities. \cite{reif2009fundamentals}
Also notable is the identity
\begin{align}
    \label{eq:chain3}
    \frac{\pslash{x}{y}{z}}{\pslash{x}{y}{w}}
    =
    \frac{\pslash{w}{z}{\smash{y}}}{\pslash{w}{z}{x}}.
\end{align}
This identity can be used for deriving
\begin{equation}
    \label{eq:cp/cv}
    \frac{C_P}{C_V} = 
    \frac{\pslash{S}{T}{P}}{\pslash{S}{T}{V}} = 
    \frac{\pslash{V}{P}{T}}{\pslash{V}{P}{S}} = 
    \frac{\kappa_T}{\kappa_S},
\end{equation}
where $\kappa_T$ and $\kappa_S$ are isothermal and isentropic compressibility, respectively. \cite{nash1964carnot}
The sunray diagram again proves to be useful.
By tracking the arrow sliding by associating the scaling factors properly, one can intuitively understand that figure \ref{fig:proof2} derives equations \eqref{eq:chain1} and \eqref{eq:chain2}.
Equation \eqref{eq:chain3} also admits a graphical proof, as shown in figure \ref{fig:proof3}.
However, for the latter case, it would be advisable to take advantage of an extra structure---the oriented area structure---which will be introduced in the next section.


There are at least two strengths of the graphical method.
First, it serves as a quick mnemonic for partial derivative identities: if the configuration of arrows corresponding to an identity is reproduced, one can quickly read out the identity from the sunray diagram.
The second is that the graphical way enables one to see the blueprint of proofs.
Students may not be sure about how to transform the left-hand side into the right-hand side of a partial derivative identity.
However, with the sunray diagram, what they should do is simply finding a pathway connecting the given initial and final arrows.
Finding such a pathway is often straightforward by graphical reasoning.
Furthermore, suppose only the left-hand side of equation \eqref{eq:chain2} is given.
Students may not be sure about how to progress into another expression.
In this case, sunray diagrams will hint possible directions to progress, allowing students to respond to various partial derivative calculations actively.

\begin{figure}[t!]
    \centering
    \includegraphics[scale=1.0]{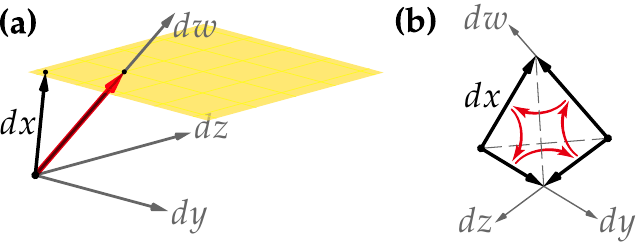}
    \caption{(a) The red arrow represents $\pslash{x}{w}{y,z} dw$: the projection of $dx$ to the direction parallel to $dw$ along the ``sunplane'' spanned by $dy$ and $dz$. The linear independence requirement is reflected by the fact that $dw$ fails to make an intersection with the ``sunplane'' if $dy$, $dz$, and $dw$ are coplanar. (b) $dx$ runs a lap around the edges of a tetrahedron that four of its edges parallel to $dx$, $dy$, $dz$, and $dw$, respectively. The resulting identity reads $\pslash{x}{y}{z,w}\pslash{y}{z}{x,w}\pslash{z}{w}{x,y}\pslash{w}{x}{y,z}=+1$.}
    \label{fig:sunplane-tetra}
\end{figure}

Before moving on to the next section, it is worth noting that our considerations are restricted to planar sunray diagrams in this paper.
That is, two-dimensional linear systems of differentials are being considered only.
When considering equations \eqref{eq:chain1}, \eqref{eq:chain2}, and \eqref{eq:chain3_re}, it is assumed that each of the four variables $x$, $y$, $z$, and $w$ can be considered as a function of two others, such as $x = f(y,z)$ and $x = g(y,w)$.
A typical beginners' question is that what becomes different when each of $x$, $y$, $z$, and $w$ can be considered as a function of the other three, instead of two.
If this was the case, we should have drawn the sunray diagrams of figures \ref{fig:proof2} and \ref{fig:proof3} in a three-dimensional space, not on a plane.
Then, the arrows are slid along a ``sunplane'' (a partial derivative fixes two variables), and again, various identities can be easily obtained by arrow sliding (figure \ref{fig:sunplane-tetra}).
Interested readers might want to explore higher-dimensional partial derivative identities by such ``sunplane'' diagrams.

\subsection{\label{sec:SunrayWedge}The Oriented Area}

For the moment, the sunray diagram method is introduced as a set of ad hoc rules associating diagrams to mathematical expressions to obtain partial derivative identities easily: just a notation change.
Although the concerns about the validity of identifying infinitesimals such as $dx$ with directed arrows (vectors) were ignored, the practical standpoint has brought fruitful results.

One more structure that can be introduced in our graphical language is the oriented area.
The lack of geometrical semantics makes the validity of its introduction indeterminate: it is not clear whether the notion of the oriented area can be physically relevant or observable, being invariant under some ``coordinate'' (or basis) transformation.
However, we again take an ad hoc or practical stance and let its physical relevance be self-evident.

\begin{figure}[t]
    \centering
    \includegraphics{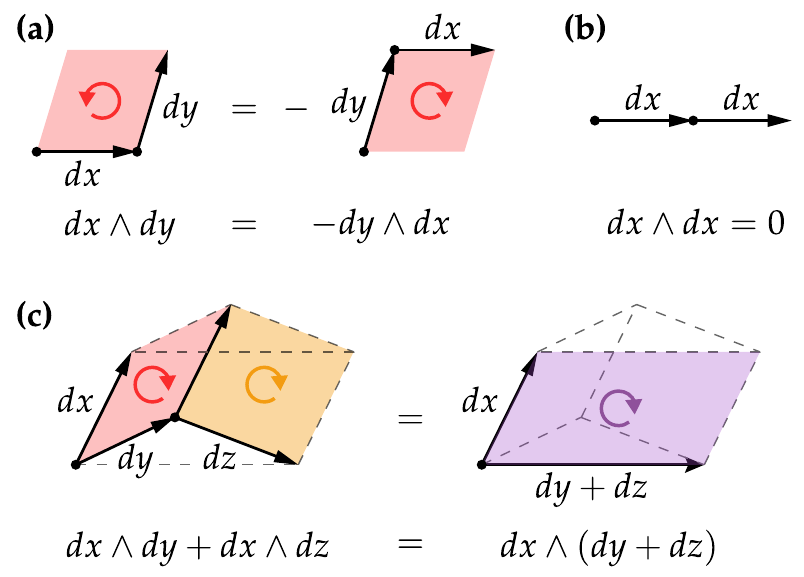}
    \caption{The properties of the wedge operation. The last diagram lies in a two-dimensional plane and should not be seen as a three-dimensional rooflike structure.}
    \label{fig:wedge-property}
\end{figure}

Given two differentials $dx$ and $dy$, a binary operation $\wedge$ (read as ``wedge'') is defined to give the area of the oriented parallelogram generated by arrows representing $dx$ and $dy$ in their sunray diagram.
Its sign is positive when the orientation of the parallelogram is anticlockwise (i.e., the thumb points upward when the right-hand is winded from $dx$ to $dy$).
Then, $dx\wedge dy = -dy \wedge dx$, which in turn implies that $dx\wedge dx = - dx\wedge dx = 0$.
Also, $\wedge$ is distributive with respect to $+$.
These fundamental properties of the wedge product are illustrated in figure \ref{fig:wedge-property}.

\begin{figure}[t]
    \centering
    \includegraphics{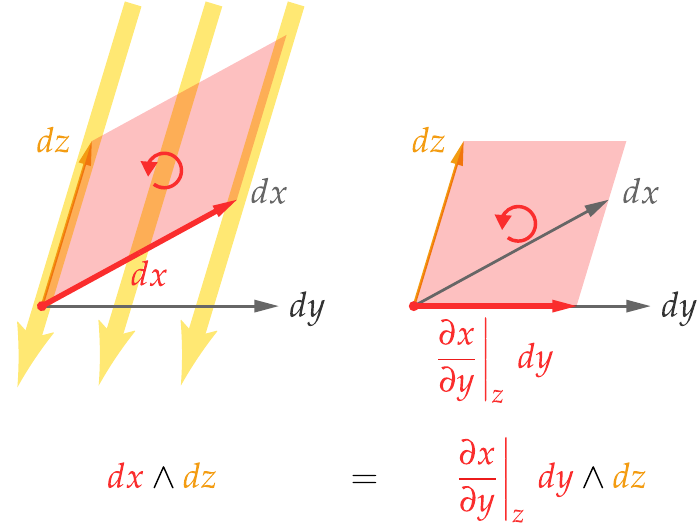}
    \caption{Accompanying oriented parallelograms are shown while the thick red arrow is slid from $dx$ to $\pslash{x}{y}{z}dy$.
    }
    \label{fig:equalarea}
\end{figure}

The introduction of the oriented area structure enables the interpretation of a partial derivative $\pslash{x}{y}{z}$ as a ratio between two geometrical quantities.
Note that the scale factor of an arrow sliding is not a ratio between ``lengths'' of starting and ending arrows.
In fact, we cannot even argue about the ``lengths'' of arrows, as there is no metric structure in sunray diagrams.
Instead, the scale factor is a ratio between two oriented areas:
\begin{equation}
    \label{eq:equalarea}
    dx \wedge dz = \pfrac{x}{y}{z} \, dy \wedge dz.
\end{equation}
Algebraically, this follows from operating ``$\wedge\,dz$'' on equation \eqref{eq:pdcoeff}.
Geometrically, its illustration is given in figure \ref{fig:equalarea}.
Note that the oriented area of a parallelogram does not change by a ``sunray sliding'' (i.e., a shear transformation) along one of its edges.
Now, one may attempt to rewrite equation \eqref{eq:equalarea} as
\begin{equation}
    \label{eq:frac}
    \pfrac{x}{y}{z} = \frac{dx \wedge dz}{dy \wedge dz},
\end{equation}
and, as it turns out later, it is much convenient to use this form when algebraically manipulating partial derivatives.
As the wedge product between two differentials is defined as a signed area, the right-hand side of equation \eqref{eq:frac} is well-defined as a division of two real numbers.
In general, such a fractional expression is well-defined only for coplanar (parallel) oriented areas, but in virtue of the two-dimensional nature of our systems, they are well-defined for every pair of oriented areas in our case.


It is instructive to re-derive the aforementioned partial derivative identities with this new apparatus, the oriented area.
For example, equation \eqref{eq:chain3} can be proved by the following trick:
\begin{align}
    \label{eq:chain3_re}
    \frac{\pslash{x}{y}{z}}{\pslash{x}{y}{w}}
    &= \frac{\left.{dx\wedge dz}\right/{dy\wedge dz}}{\left.{dx\wedge dw}\right/{dy\wedge dw}} \nonumber
    \\
    &= \frac{\left.{dy\wedge dw}\right/{dy\wedge dz}}{\left.{dx\wedge dw}\right/{dx\wedge dz}}
    = \frac{\pslash{w}{z}{\smash{y}}}{\pslash{w}{z}{x}}.
\end{align}
A corresponding parallelogram-based visualization is also possible and straightforward, but as expected, it appears to be a bit complicated than the vector-based visualization, figure \ref{fig:proof3}.

Interpreting partial derivatives as ratios between two oriented areas and doing the ``wedge gymnastics'' is identical to the technique of Jacobian that has been employed in the thermodynamics literature in essence.\cite{crawford1949jacobian,carroll1965use,landau1980sp}
From geometrical interpretation or by several algebraic manipulations, it can be shown that the ratio between two oriented areas $dx\wedge dy$ and $dz \wedge dw$ equals
\begin{equation}
    \frac{dx\wedge dy}{dz\wedge dw}=\pfrac{x}{z}{w}\pfrac{y}{w}{z}-\pfrac{y}{z}{w}\pfrac{x}{w}{z},
\end{equation}
which is the Jacobian ${\partial(x,y)}/{\partial(z,w)}$.
Thus, the Jacobian technique can be translated into the visual calculus of the oriented area and vice versa.

Also note that, although the wedge operation here is defined intrinsically in terms of sunray diagrams, its algebra is isomorphic to the conventional notion of the wedge operation on multivectors and multiforms (refer to pp. 117-126 of \cite{schutz1980geometrical}).
The connection between the two will be self-evident later.

\subsection{\label{sec:SunrayApplications}Application to Thermodynamics}

Now, we demonstrate the real use of sunray diagrams in deriving thermodynamic identities.
Our example is equation \eqref{eq:cp-cv}, which is the unproven one among the three examples given earlier, equations \eqref{eq:thermprop}, \eqref{eq:cp-cv}, and \eqref{eq:cp/cv}.
Other various thermodynamic identities can be readily worked out by case by case applications of the techniques introduced in this paper.

\begin{figure}[t]
    \centering
    \includegraphics{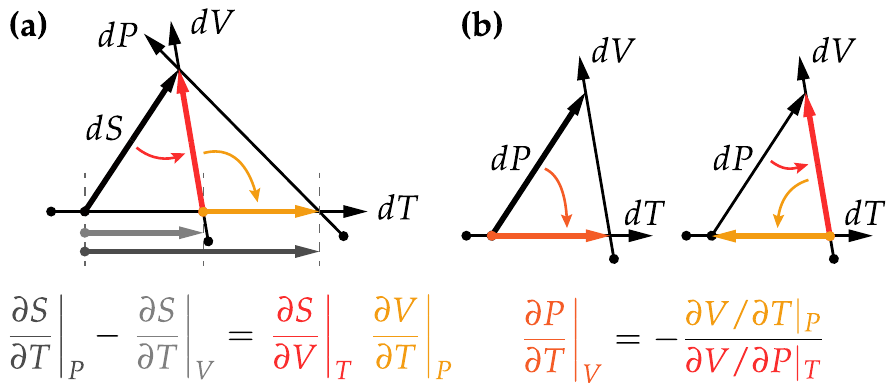}
    \caption{Two of the sunray diagrams used when deriving equation \eqref{eq:cp-cv}. 
    Since identities to be obtained from these diagrams are rather ``mathematical'' ones, the condition $dP \wedge dV = dT \wedge dS$ is ignored when drawing them.}
    \label{fig:cp-cv}
\end{figure}

For simplicity, a single-component system that has two thermodynamic degrees of freedom with its particle number fixed is assumed.
Then, the differentials $dP$, $dV$, $dT$, and $dS$ are the inhabitants of our sunray diagram. From the definition of heat capacities,
\begin{equation}
    \label{eq:der1}
    C_P - C_V = T \pfrac{S}{T}{P} - T \pfrac{S}{T}{V}.
\end{equation}
For the right-hand side of equation \eqref{eq:der1}, a sunray diagram in figure \ref{fig:cp-cv}(a) is drawn to give
\begin{equation}
    \label{eq:der2}
    \pfrac{S}{T}{P} - \pfrac{S}{T}{V} = \pfrac{S}{V}{T} \, \pfrac{V}{T}{P}.
\end{equation}
To transform $\pslash{S}{V}{T}$ in equation \eqref{eq:der2} into a more tractable form, use one of Maxwell relations,
\begin{equation}
    \label{eq:SVPT}
    \pfrac{S}{V}{T} = \pfrac{P}{T}{V},
\end{equation}
then draw a sunray diagram in figure \ref{fig:cp-cv}(b) to obtain $\pslash{P}{T}{V} = - ({\pslash{V}{T}{P}}) \, / \, ({\pslash{V}{P}{T}})$. 
Finally, equation \eqref{eq:cp-cv} is derived, identifying $\kappa_T$ with $-(1/V) \, \pslash{V}{P}{T}$ and $\alpha_P$ with $(1/V) \, \pslash{V}{T}{P}$.

\begin{figure}[t]
    \centering
    \includegraphics{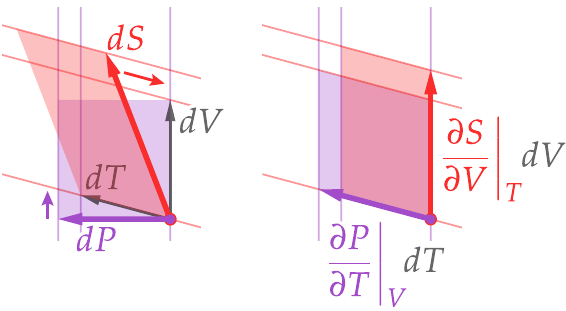}
    \caption{The sunray diagram that derives a Maxwell relation, equation \eqref{eq:SVPT}. 
    The orientation of all the parallelograms in this figure is clockwise.
    The parallelogram $dT\wedge dS$ is slid to be  $\pslash{S}{V}{T} dT\wedge dV$ (colored in red).
    At the same time, $dP\wedge dV$ is slid to be  $\pslash{P}{T}{V} dT\wedge dV$ (colored in purple).
    Since $dP\wedge dV = dT\wedge dS$, these two areas are the same, and equation \eqref{eq:SVPT} is proved.
    }
    \label{fig:maxray}
\end{figure}

A Maxwell relation equation \eqref{eq:SVPT}, also deserves a graphical counterpart.
It boils down to an additional identity,
\begin{equation}
    \label{eq:d1stlaw}
    dT \wedge dS = dP \wedge dV.
\end{equation}
Provided this, all of Maxwell relations can be derived in a remarkably simple manner: e.g.,
\begin{align}
    \label{eq:der3}
    \pfrac{S}{V}{T} &= \frac{dT \wedge dS}{dT \wedge dV} = \frac{dP \wedge dV}{dT \wedge dV} 
    = \pfrac{P}{T}{V}.
\end{align}
Figure \ref{fig:maxray} shows a diagrammatic representation;
equating the scale factors of sliding oriented areas $dP \wedge dV$ and $dT \wedge dS$ to a common parallelogram $dT\wedge dV$ leads to $\pslash{P}{T}{V}=\pslash{S}{V}{T}$.

Strictly speaking, equation \eqref{eq:d1stlaw} is a relation about the oriented areas of parallelograms formed by differentials $(dT,dS)$ and $(dP,dV)$ in a sunray diagram by definition.
However, our previous knowledge of thermodynamics strongly suggests to ``confuse'' it with the relation about infinitesimal area elements in $P$-$V$ and $T$-$S$ graphs---recall that the first law of thermodynamics implies that the area of a particular cycle is the same when the cycle is plotted on $P$-$V$ and $T$-$S$ graphs.
This is the point where the mathematical syntax meets physical semantics.
If we interpret oriented areas in sunray diagrams as infinitesimal area elements in thermodynamic state space, it will be concluded that \textit{a sunray diagram depicts a zoom-in of an infinitesimal region in the thermodynamic state space}.

Still, there is a room for clarifications, as the physical meaning of ``arrows'' or ``vector differentials'' in sunray diagrams remains mysterious.
Indeed, differentials such as $dP$ or $dV$ do not carry a ``direction'': they are just numbers, albeit infinitesimally small.
Meanwhile, $dP$, in some sense, suggests an image of ``movement in the $P$-direction,'' and it seems that it might be possible to recast it as a directed quantity.
A hint may be found from Maxwell's geometrical interpretation of thermodynamic partial derivative identities,\cite{maxwell1891theory} where a similar notion of ``directed differential'' arises as a ``unit increment'' of thermodynamic variables.

\section{\label{sec:Exp}The Conceptual Basis for the Sunray Diagram}

Now, we turn up the math level a notch and establish the conceptual basis for the sunray diagram in the language of differential geometry.
Elementary knowledge of differential geometric notions is assumed, sections 5.1--5.4 of Nakahara \cite{nakahara2003geometry} or equivalent.
For the calculus of differential forms (the wedge operation and the exterior derivative) and its applications to thermodynamics, refer to pp 113--167 of Schutz. \cite{schutz1980geometrical}
Also, this section visualizes multivectors and differential forms following Schutz \cite{schutz1980geometrical} and Misner, Thorne, Wheeler. \cite{mtw2000gravitation}
To be self-contained, we provided an explanation of the visualization rules for multivectors and differential forms in the appendix.
Meanwhile, notations that somewhat deviate from the standard is used to emphasize the correspondence between the algebra of vectors and one-forms as well as to distinguish ordinary differentials from differential forms (cf. the syntax-semantics discussion at the end of section \cref{sec:ExpSunray}).
We denote vectors with right-headed arrows ($\vec{v}$), one-forms with left-headed arrows ($\lvec{w}$), and their contraction by juxtaposition ($\lvec{w}\vec{v}$).
The exterior derivative is denoted by $\d$, and a two-form is denoted with boldface ($\bm{\gamma}$).
The relationship with the standard notations is explained in the appendix.
Lastly, we note that although group-theoretical jargons are used to clarify concepts in section \cref{sec:ExpCovariance}, a moderate level of understanding of symmetries in physics will suffice to catch the essence.

\subsection{The ``Covariance'' of Sunray Diagrams\label{sec:ExpCovariance}}
The discussions so far can be rephrased in mathematical terms as a successive reduction of the invariance group.
In the first section, the sunray diagram is introduced as a visualization of differentials in thermodynamics as vectors.
Recall that, being closed under addition and scalar multiplication, a system of differentials forms a vector space.
Neglecting physical content, a vector in a sunray diagram is merely a schematic visualization of an element of such a vector space: ``vectors are elements of a vector space.''
Then, partial derivatives come from linear algebraic relationships between the vectors of a linear system of differentials.

However, such ``minimal'' linear algebraic conceptualization of vectors often fails to be relevant or useful in the physics literature.
In our case, major problems are twofold.
First, the definition ``vectors are elements of a vector space'' is too abstract or axiomatic and far from making contact with physical objects.
What does the ``vector'' $dP$ in a sunray diagram have to do with the value of infinitesimal pressure increment or a vertical or horizontal line segment in the $P$-$V$ plane?
Second, in physics, there is a notion of observer (or frame) independence, and often vectors are meant to be an invariant object with respect to a given invariance group.
For example, in three-dimensional Euclidean vector calculus, vectors are one of the ``data types'' that are invariant under three-dimensional rotations of the basis vectors, while their components are contravariant.

These two statements boil down to the discrepancy between mathematicians' linear algebraic concept of vectors and physicists' concept of vectors that emphasizes the physical significance of invariance groups that connects between frames.
In terms of group theory,\cite{sternberg1995group,tung1985group,cvitanovic2009group} the physicists' definition is that vectors are elements of the carrier space $\mathrm{X}$ of a representation $\pi:\mathrm{G}\rightarrow\Aut(\mathrm{X})$ of the invariance group $\mathrm{G}$, usually the fundamental representation.\footnote{
By fundamental representation of a group we mean that the smallest, faithful representation of the group, following the convention used in the physics literature.\cite{tung1985group,cvitanovic2009group}}
In other words, vectors are the elements of a vector space with a group action preserving the linear algebraic structure defined in it.
This definition is precisely the idea behind the catchphrase ``vectors are objects that transform like a vector,'' as put by Griffiths.\cite{griffiths2012em}
Then, tensors are obtained from tensor representations that take their carrier spaces as tensor products of multiple copies of the vector representation carrier space.
In this perspective of view, the mathematicians' vectors are the special case where the invariance group is the general linear group, $\mathrm{GL}(n)$.
In contrast, physicists often assume more restrictive invariance groups and pay more attention to the invariance group than the linear structure for the physical semantics.
For instance, recall the earlier example of vectors in three-dimensional Euclidean vector calculus.
Such ``$\mathrm{SO}(3)$-vectors,'' i.e., quantities that carry the fundamental representation of the three-dimensional special orthogonal group $\mathrm{SO}(3)$, differ from the ``$\mathrm{GL}(3)$-vectors,'' the elements of the three-dimensional group-reacting vector space equipped with the largest three-dimensional invariance group.

In the sunray diagram case, the invariance group establishes the equivalence between diagrams (i.e., how a diagram appears in different ``frames,'' e.g., $P$-$V$ and $T$-$S$ graphs) and determines what quantities are geometrically or physically meaningful.
Up to section \ref{sec:SunrayBasic}, the best one can conclude is that the invariance group of sunray diagrams is a subgroup of the general linear group $\mathrm{GL}(2)$.
An arbitrary $\mathrm{GL}(2)$ transformation of a sunray diagram does not affect the partial derivative identity it represents.
The mathematicians' minimal notion of vectors is not challenged due to the absence of a hint about the invariance group.

Then, section \ref{sec:SunrayWedge} introduces the oriented area and finds its rationale from the usefulness of oriented areas, namely, the ratio interpretation of partial derivatives and the wedge gymnastics.
Still, the invariance group remains at $\mathrm{GL}(2)$ if the oriented areas themselves are not physically meaningful quantities but only their ratios, Jacobians, are significant.
In this case, the oriented areas should appear only in transient steps when working with physical quantities written in partial derivatives.
However, contact with physical semantics in section \ref{sec:SunrayApplications} implies that this is not the case.
It suggests interpreting the oriented areas as infinitesimal area elements in thermodynamic state space, which are physically relevant quantities, namely, infinitesimal work.
The preservation of oriented areas under basis transformations finally requires the invariance group to be reduced to $\mathrm{SL}(2)$, the two-dimensional special linear group, which is isomorphic to $\mathrm{Sp}(2)$, the two-dimensional symplectic group.\footnote{In general, an inspection on the first law of thermodynamics shows that the invariance group is $\mathrm{Sp}(2n)$ for sunray diagrams of $2n$-dimensional thermodynamic state space.}

\subsection{Maxwell's Method as a Graphical Calculus of Differential Forms\label{sec:ExpMaxwell}}
The invariance group line of thought revealed that the ``vector differentials'' in planar sunray diagrams carry a representation of the two-dimensional symplectic group $\mathrm{Sp}(2)$.
Moreover, another lesson from section \ref{sec:SunrayApplications} hints that a sunray diagram depicts a local neighborhood of a point in thermodynamic state space.
These properties strongly hint that the sunray diagram is a representation of some differential geometric structure.
An interpretation that fulfills both the properties is that the ``vector differentials'' of sunray diagrams live in a tangent or cotangent space of the thermodynamic state space, while we expect that the tangent space is the answer to reproduce the vector nature of sunray diagrams.

The last missing step is to make contact with the familiar geometric notions such as ordinary differentials or infinitesimal displacement and explicate the physical or numerical content of the ``vector differentials.''
The final objective will be locating (overlaying) a sunray diagram in the familiar $P$-$V$ or $T$-$S$ graphs.
A good reference point is Maxwell's approach, as it has a direct physical interpretation in the thermodynamic state space and, at the same time, can be recast into the language of differential forms.

\begin{figure}[t]
    \centering
    \includegraphics{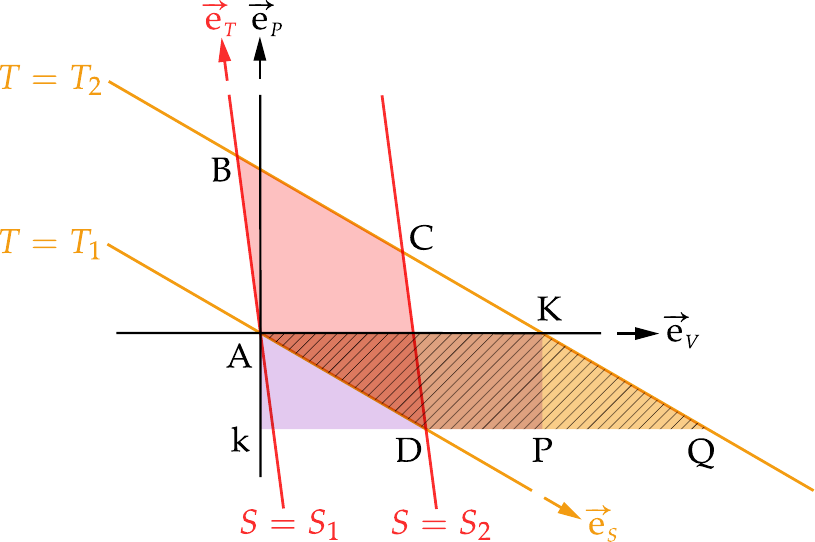}
    \caption{The diagram for Maxwell's graphical method.}
    \label{fig:maxwell}
\end{figure}

Maxwell incorporated a graphical method when he derived the relations that are now well-known as Maxwell relations in his seminal work on thermodynamics.\cite{maxwell1891theory}
He started by considering the diagram shown in figure \ref{fig:maxwell}.
Two isothermal lines and two adiabatic lines are overlaid on a $P$-$V$ plot.
Each corresponds to a small difference of temperature and entropy (i.e., $\abs{T_2-T_1}\ll T_1$ and $\abs{S_2-S_1}\ll S_1$) so that they appear in straight lines in the figure.
It is customary to set $T_2-T_1$ and $S_2-S_1$ to be unit temperature and entropy, with the understanding of the local linearization; let us adopt this convention, following Maxwell.
The area of the parallelogram ABCD is $(T_2-T_1)(S_2-S_1)=(1\text{ unit of energy})$ since a unit area in the $T$-$S$ plane appears as a unit area in the $P$-$V$ plane.

The areas of parallelograms ABCD and AKQD are equal, as both are between the same parallel lines with the same baseline $\overline{\text{AD}}$.
Similarly, the parallelograms AKQD and AKPk are also equal in their area.
Hence, 
\begin{equation}
    \label{eq:maxwellderiv1}
    \overline{\text{AK}} \cdot \overline{\text{Ak}} = \left\vert \text{AKPk} \right\vert = \left\vert \text{ABCD} \right\vert = (1\text{ unit of energy}).
\end{equation}
Then, Maxwell interprets the physical meaning of the lengths of the segments.
$\overline{\text{AK}}$ corresponds to the increased volume for a unit increase of temperature while the pressure is held constant.
Likewise, $\overline{\text{Ak}}$ is the decreased pressure for a unit increase of entropy while the temperature is held constant.
If we interpret this in terms of partial differentials, equation \eqref{eq:maxwellderiv1} translates to
\begin{equation}
    \label{eq:maxwellderiv2}
    - \pfrac{V}{T}{P}\cdot \pfrac{P}{S}{T} = 1,
\end{equation}
which the reciprocity relation equation \eqref{eq:reci} can be applied to give a Maxwell relation,
\begin{equation}
    \label{eq:maxwellderiv3}
    \pfrac{V}{T}{P} = -\pfrac{S}{P}{T}.
\end{equation}
The other three Maxwell relations can also be derived by applying the same procedure with different equal-area slidings instead of $\overline{\text{AK}} \cdot \overline{\text{Ak}}$.

What we unearth from the graphical procedure of Maxwell is the calculus of differential forms.
Maxwell's parallelogram ABCD is a linearization of the curved quadrangle bounded by a pair of nearby isothermal and adiabatic lines.
Thus, Maxwell's diagram can be interpreted as lying on the tangent space to the thermodynamic state space at point A, while the unit-spaced isothermal and adiabatic contour lines being the visualizations of differential one-forms $\d T$ and $\d S$ at A.
Respectively, the oriented area of the parallelogram ABCD visualizes the differential two-form $\d T \wedge \d S$.

\begin{figure}[t]
    \centering
    \includegraphics{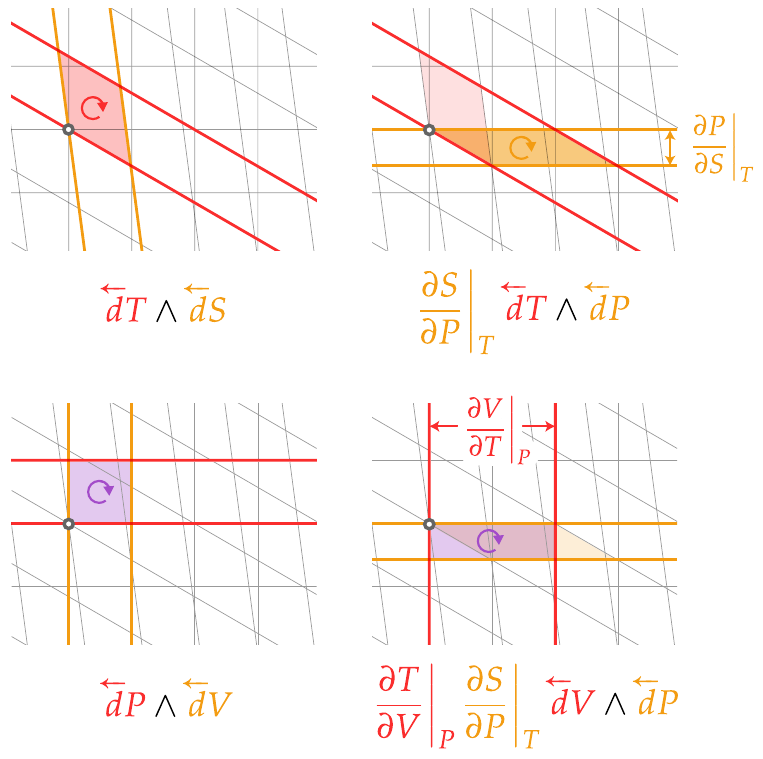}
    \caption{Transformations between two-forms. All are representing the same two-form, $\d T\wedge \d S = \d P\wedge \d V$.}
    \label{fig:maxwell-form}
\end{figure}

Next, Maxwell's geometric interpretation of $\overline{\mathrm{AK}}$ as $\pslash{V}{T}{P}$ means that the set of contour lines  $\overline{\mathrm{Ak}}$ and $\overline{\mathrm{KP}}$ correspond to a one-form $(\pslash{V}{T}{P})^{-1}\, \d V = \pslash{T}{V}{P}\, \d V$.
The reciprocal scaling is due to the nature of one-forms that act on a vector of some magnitude to give a scalar: the smaller the spacing of contour lines, the larger the magnitude of the one-form.
Similarly, the parallel lines $\overline{\mathrm{AK}}$ and $\overline{\mathrm{kP}}$ correspond to the one-form $-\pslash{S}{P}{T}\, \d P$.
When these two one-forms are wedge producted, what is obtained is a two-form that has its unit cell AKPk, which equals to a two-form of ABCD, $\d T\wedge\d S$, according to the equal-area sliding.
In sum, Maxwell's equal-area sliding boils down to expressing the same two-form $\d T \wedge \d S$ in various bases:
\begin{align}
    \label{eq:maxwellderivpd}
    \d T \wedge \d S 
    &= \pfrac{S}{P}{T} \, \d T \wedge \d P \nonumber \\
    &= \pfrac{T}{V}{P} \pfrac{S}{P}{T} \, \d V \wedge \d P.
\end{align}
Figure \ref{fig:maxwell-form} shows how the algebraic steps of equation \eqref{eq:maxwellderivpd} correspond to successive shear transformations of an egg-crate.
One can observe that the movement of the unit cell of the egg-crate exactly matches with Maxwell's parallelogram sliding.

Finally, from the first law of thermodynamics (and the integrability of infinitesimal heat),
\begin{equation}
    \d E = T\d S - P\d V
    \implies
    \d T\wedge\d S = \d P\wedge\d V.
    \label{eq:d1stlaw_form}
\end{equation}
Note that $\d\,\d E = 0$.
This serves as a mathematical basis for the ``infinitesimal cycle in $P$-$V$ and $T$-$S$ graphs'' argument for justifying equation \eqref{eq:d1stlaw} that appealed to physical intuition.
Provided this, equation \eqref{eq:maxwellderiv3} can be derived from equation \eqref{eq:maxwellderivpd} by
\begin{align}
    (-1) \, \d V \wedge \d P &= \d P \wedge \d V = \d T \wedge \d S \nonumber \\
    &= \pfrac{T}{V}{P} \pfrac{S}{P}{T} \, \d V \wedge \d P.
\end{align}

To sum up, Maxwell's method is a graphical calculus of differential forms visualized in a tangent space of the thermodynamic state space.
The ``geometric part'' (equation \eqref{eq:maxwellderivpd}) that utilizes egg-crate sliding boils down to the following arithmetic of differential forms:
\begin{align}
    \label{eq:pdcoeff_form}
    \d x = \pfrac{x}{y}{z} \, \d y + \pfrac{x}{z}{y} \, \d z;
    \\
    \label{eq:equalarea_form}
    \d x \wedge \d z = \pfrac{x}{y}{z} \, \d y \wedge \d z.
\end{align}
The ``thermodynamic part'' (equation \eqref{eq:d1stlaw_form}) refers to a physical law and demands that the $P$-$V$ and $T$-$S$ planes are related by a symplectic transformation, i.e., an area-preserving diffeomorphism.

\subsection{The Sunray Diagram as a Graphical Calculus of Symplectic Gradients\label{sec:ExpSunray}}
Maxwell's graphical method is a ``form implementation'' of the syntax of differentials, equations \eqref{eq:pdcoeff}, \eqref{eq:equalarea}, and \eqref{eq:d1stlaw}: equations \eqref{eq:pdcoeff_form}, \eqref{eq:equalarea_form}, and \eqref{eq:d1stlaw_form}.
Then, it is natural to ask whether a ``vector implementation'' is possible.
In fact, the symplectic structure provides a dual map between one-forms and vectors so that the world of differential forms can be translated into ``vector differentials.''
We then anticipate that the sunray diagram arises from the vector picture.

\begin{figure}[t]
    \centering
    \includegraphics{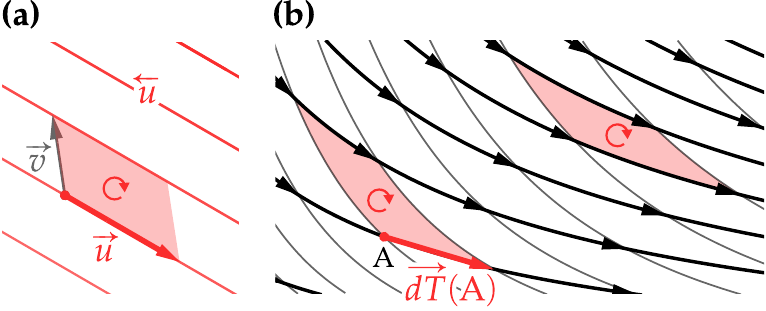}
    \caption{(a) According to equation \eqref{eq:uvone}, given a one-form $\lvec{u}$ (depicted by red contour lines), one can construct its symplectic dual, $\vec{u}$, by ``squeezing'' a unit oriented area between two consecutive contour lines then taking its baseline.
    The direction of $\vec{u}$ must conform with the orientation of the squeezed area, as shown in the figure.
    (b) Locating a sunray diagram on a $P$-$V$ graph.
    The figure shows a region in the $P$-$V$ plane.
    The symplectic gradient vector field $\longvec{dT}$ (black arrows) flows along the isotherms and increases the value of entropy by one unit.
    The areas of cells formed by the unit-spaced isothermal (thick) and adiabatic (fine) lines are always the same as the unit area.
    If one picks symplectic gradient vectors at a particular point and performs the arrow sliding, a sunray diagram is obtained.}
    \label{fig:gradvec}
\end{figure}

Equation \eqref{eq:d1stlaw_form} can be understood as stating the invariance of the symplectic form
\begin{equation}
    \bm{\omega} := \d P \wedge \d V = \d T \wedge \d S,
\end{equation}
under the coordinate transformation between $(P,V)$ and $(T,S)$.\footnote{
The contact geometry formulation of thermodynamics \cite{hermann1973geometry,mrugala1978geometrical,bravetti2015contact} sets the symplectic form as $\d(-\d E + T \d S - P \d V ) = \d T\wedge \d S - \d P \wedge \d V$.
In this approach, $\d T \wedge \d S$ can be effectively identified with $\d P \wedge \d V$ only on a Lagrangian surface in a Legendrian submanifold.
However, we take an alternative view that is more well-suited to our scope, equilibrium states of two thermodynamic degrees of freedom.
The thermodynamic state space is identified with a two-dimensional symplectic manifold $\mathcal{M}$ equipped with the symplectic form $\d P \wedge \d V = \d T \wedge \d S$, and $(P,V)$ and $(T,S)$ are regarded as two different coordinate charts on $\mathcal{M}$ that are related to each other by a canonical transformation (symplectomorphism).
One can further confirm that the two approaches are consistent with each other by considering the equation of state and the Gibbs-Duhem relation.
Refer to Kocik \cite{kocik1986geometry} also.
}
Being coordinate-invariant, $\bm{\omega}$ has a qualification to be a physically relevant object, and in fact, it is already granted with such position as work of infinitesimal cycles.
Resultingly, the ``symplectic dual'' map, defined as the map sending a one-form $\lvec{u}$ to the vector $\vec{u}$ satisfying 
\begin{equation}
    \label{eq:symplecticdual}
    \left<\lvec{u},\vec{v}\right>= -\left<\bm{\omega},\vec{u}\wedge \vec{v}\right>
\end{equation}
for arbitrary vectors $\vec{v}$, is coordinate-invariant.
This dual map is invertible, provided that $\bm{\omega}$ is nondegenerate.
In graphical terms, $\vec{u}$ is obtained by ``squeezing'' a unit oriented area between two nearby contour lines that visualize $\lvec{u}$: see figure \ref{fig:gradvec}(a).
This can be verified by checking the direction and magnitude of $\vec{u}$.
It is parallel to the contour lines depicting $\lvec{u}$ because $\left<\lvec{u},\vec{u}\right>=-\left<\bm{\omega},\vec{u}\wedge\vec{u}\right>=-\left<\bm{\omega},0\right>=0$; for a vector $\vec{v}$ that pierces $\lvec{u}$ once, $\vec{v}\wedge\vec{u}$ is a bivector of unit area: 
\begin{equation}
    \left<\lvec{u},\vec{v}\right>=1 \implies
    \left<\bm{\omega},\vec{v}\wedge \vec{u}\right> = \left<\lvec{u},\vec{v}\right> =1.
    \label{eq:uvone}
\end{equation}

Then, the ``symplectic gradient''\cite{schutz1980geometrical,symplectic1,symplectic2,cariglia2013kepler} $\longvec{dA}$ of a scalar field $A$ is defined as the symplectic dual of $\d A$:
\begin{equation}
    \label{eq:symplecticdual_form}
    \left<\d A,\vec{v}\right> = -\left<\bm{\omega},\longvec{dA}\wedge\vec{v} \right>
\end{equation}
for every vector field $\vec{v}$.
In component terms, in the $(P,V)$ coordinate system for instance,
\begin{equation}
    \label{eq:symplecticdual_form_comp_PV}
    \longvec{dA}
    =
    +\pfrac{A}{P}{V} \vec{\mathrm{e}}_V
    -\pfrac{A}{V}{P} \vec{\mathrm{e}}_P.
\end{equation}
This can be confirmed by putting $\longvec{dA}
=
(dA)^{\hspace{-0.10em}P}
\hspace{0.07em}\vec{\mathrm{e}}_P
+
(dA)^{\hspace{-0.15em}V} 
\hspace{0.05em}\vec{\mathrm{e}}_V$
and $\bm{\omega}=\d P \wedge \d V$ into equation \eqref{eq:symplecticdual_form}, where $\vec{\mathrm{e}}_P$ and $\vec{\mathrm{e}}_V$ denote the coordinate basis vectors of the $(P,V)$ coordinate system.
In visual terms, the symplectic gradient vector field $\longvec{dA}$ flows along the contour lines of $A$: see figure \ref{fig:gradvec}(b).
It is the ``movement along the constant-$A$ direction'' and differs from what is called the ``gradient vector'' $\nabla A$ in Euclidean spaces.\footnote{Note that, in the absence of a metric structure, a gradient vector cannot exist, as its definition involves an inner product.
In layman's terms, a measure of length is required to define the direction ``orthogonal'' to contour lines.}

The physical meaning of the symplectic gradient vector field $\longvec{dA}$ is a flow that transports a point on the thermodynamic state space to a position that the value of $B$ increased by one unit, where $A$ is a conjugate variable of $B$.
By $A$ being conjugate to $B$, it means that $\d (A\d B) =\d A\wedge \d B = \bm{\omega}$, i.e., the Jacobian $\partial(A,B)/\partial(P,V)$ is equal to one (then, $-B$ is conjugate to $A$).\footnote{In the symplectic formulation of classical mechanics, the symplectic gradient $\longvec{dA}$ coincides with the flow generated by $A$, $\{\,\,\,,A\}$, where $\{\,\,\,,\,\,\,\}$ is the Poisson bracket. E.g., a momentum $p$ generates displacements of its conjugate position $q$: $\{q,p\}=\longvec{dp}[q] = \vec{\mathrm{e}}_q[q]= 1$.}
For example, $\longvec{dP}=+\vec{\mathrm{e}}_{V}$ and $\longvec{dV}=-\vec{\mathrm{e}}_{P}$.
Similarly, $\longvec{dT}=+\vec{\mathrm{e}}_{S}$ and $\longvec{dS}=-\vec{\mathrm{e}}_{T}$.

Now, taking the symplectic dual to equation \eqref{eq:pdcoeff_form} then taking vector wedge product with $\longvec{dz}$ yields
\begin{align}
    \label{eq:pdcoeff_vec}
    \longvec{dx} = \pfrac{x}{y}{z} \, \longvec{dy} + \pfrac{x}{z}{y} \, \longvec{dz};\\
    \label{eq:equalarea_vec}
    \longvec{dx} \wedge \longvec{dz} = \pfrac{x}{y}{z} \, \longvec{dy} \wedge \longvec{dz}.
\end{align}
It is also straightforward to verify that
\begin{equation}
    \label{eq:d1stlaw_vec}
    \longvec{dP} \wedge \longvec{dV} = \longvec{dT} \wedge \longvec{dS}.
\end{equation}
Since $\longvec{dP}$ and $\longvec{dV}$ increases $V$ and $-P$ by one unit, this quantity can be interpreted as the unit area bivector ($\longvec{dP} \wedge \longvec{dV} = \vec{\mathrm{e}}_V \wedge(-\vec{\mathrm{e}}_P) = \vec{\mathrm{e}}_P\wedge\vec{\mathrm{e}}_V$).
Equations \eqref{eq:pdcoeff_vec}, \eqref{eq:equalarea_vec}, and \eqref{eq:d1stlaw_vec} serve as a realization of the partial derivative syntax as well as the differential forms language introduced in section \ref{sec:ExpMaxwell}.

The symplectic gradient vectors are visualized as arrows in a tangent space of the thermodynamic state space, i.e., a zoom-in of the thermodynamic state space at a point.
Invariance property under coordinate change confirms that the symplectic gradient vectors are $\mathrm{Sp}(2)$-vectors.
Also, the symplectic gradient vectors can be wedge producted to form bivectors, e.g., the unit area element in equation \eqref{eq:d1stlaw_vec}.
These properties coincide with the desired properties of the ``vector differentials'' of sunray diagrams; there is no reason to hesitate to identify $\longvec{dx}$ with the arrow denoting $dx$ in sunray diagrams.
The complete picture is given in figure \ref{fig:gradvec}(b): a sunray diagram sits on a particular point of the thermodynamic state space, and an arrow $dx$ in the diagram is just the unit displacement vector along the constant-$x$ line.
By ``unit,'' it means that a conjugate variable of $x$ changes by a unit.
The arrows are not just schematic visualizations of the abstract vectors of linear algebra but are literally arrows, i.e., directed segments on (the tangent plane of) $P$-$V$ or $T$-$S$ graphs.
Provided this demystification, the sunray diagram is no more abstract or ad-hoc than Maxwell's method but has the same level of concreteness and physical content.

\begin{figure}[t!]
    \centering
    \begin{tikzpicture}
        \coordinate (xshift) at (2.2,0);
        \coordinate (yshift) at (0,-0.75);
        \coordinate (ushift) at (0,+0.75);
        \coordinate (Yshift) at (0,-2.6);
        \node (O) at (0,0) [] {$\displaystyle{dx} = \pfrac{x}{y}{z} \, {dy} + \pfrac{x}{z}{y} \, {dz}$};
        \node (O') at ($(O)+(yshift)$) [] {$dT \wedge dS = dP \wedge dV$};
        \node (Ou) at ($(O)+(ushift)$) [] {[I]};
        \node (A) at ($(O)-(xshift)+(Yshift)$) [] {$\displaystyle{{\d x} = \pfrac{x}{y}{z} \, {\d y} + \pfrac{x}{z}{y} \, {\d z}}$};
        \node (A') at ($(A)+(yshift)$) [] {$\d T \wedge \d S = \d P \wedge \d V$};
        \node (Au) at ($(A)+(ushift)$) [] {[II]};
        \node (B) at ($(O)+(xshift)+(Yshift)$) [] {$\displaystyle\longvec{dx} = \pfrac{x}{y}{z} \, \longvec{dy} + \pfrac{x}{z}{y} \, \longvec{dz}$};
        \node (B') at ($(B)+(yshift)$) [] {$\longvec{dT} \wedge \longvec{dS} = \longvec{dP} \wedge \longvec{dV}$};
        \node (Bu) at ($(B)+(ushift)$) [] {[III]};
        \node (O'') at ($(O')-(0,0.23)$) [] {};
        \coordinate (Onudge) at (0.05,0);
        \coordinate (ABnudge) at (0,0.26);
        \draw[<-,line width=0.7pt] ($(O'')-(Onudge)$) to[out=-120,in=90] ($(Au)+(ABnudge)$);
        \draw[<-,line width=0.7pt] ($(O'')+(Onudge)$) to[out=-60,in=90] ($(Bu)+(ABnudge)$);
    \end{tikzpicture}
    \caption{Two implementations of the partial derivative syntax by differential forms and symplectic gradient vectors.
    When the arrows $\lvec{}$ and $\vec{}$ of systems [II] and [III] are ``integrated out'' (ignored), both of them flows to the system [I].
    }
    \label{fig:twoimplementations}
\end{figure}

Figure \ref{fig:twoimplementations} captures the big picture of the development of this paper from the perspective of mathematical languages.
At first, the syntax of differentials and partial derivatives was introduced in an ordinary fashion, without further mathematical elaboration.
Then, the graphical notation for it, the sunray diagram, was introduced in an ad hoc manner and soon was augmented with an additional operation $\wedge$ defined on the graphical level.
Call this system ``system [I].''
However, why such prescriptions work to give valid thermodynamic partial derivative identities was unclear as well as their physical interpretation: it lacked semantics.
In search of ``microscopic realizations'' (down-to-earth constructions) of system [I], the differential forms language (system [II]) and its visual counterpart, Maxwell's graphical method, served as a good reference point.
Finally, system [III] was found to be another implementation of the system [I], and its precise semantics as symplectic gradient vectors demystified the meaning of the sunray diagram and established its mathematical and physical validity.
As a result, two semantically rich systems [II] and [III] are obtained as realizations of an unrefined, ``effective language'' (system [I]).\footnote{This scenario directly reflects the path of the authors.
When the sunray diagram method was first devised by the first author in 2014, it started as a set of graphical syntax from a practical standpoint (system [I]).
Later, its definite geometric semantics (system [III]) became apparent throughout the discussions with the second author.}

\section{\label{sec:Conclusion}Conclusion}
The sunray diagram technique provides an intuitive and handy graphical gadget for handling the partial derivative identities.
The framework enables intuitive manipulation and visualization of partial derivatives and differentials while retaining their geometric and physical meanings as symplectic gradient vectors.
Also, endowed with the constraint from the first law of thermodynamics, sunray diagrams have been shown to perform all the graphical proofs of partial derivative identities in thermodynamics.
The physical interpretation of a directed differential (such as $\longvec{dT}$) in a sunray diagram is the vector at a certain point $\text{A}$ in the thermodynamic state space that increases the value of the conjugate variable ($S$ for $\longvec{dT}$) by one unit when an infinitesimal neighborhood of $\text{A}$ is zoomed in (i.e., linearized). \footnote{In other words, the plane in which a sunray diagram is drawn is a fiber of the tangent bundle \cite{mathoriented1, mathoriented2} of the thermodynamic state space, and the directed differentials are symplectic gradient vector fields at the base point of the fiber.}
In this article, we have specialized to two-dimensional thermodynamic state spaces and formulated them as $P$-$V$ or $T$-$S$ planes; however, the interpretation of sunray diagrams with symplectic gradient vectors does not lose meaning when different formulations or higher-dimensional generalizations are considered.

The sunray diagram method can be considered as a successful reincarnation of graphical approaches of Maxwell.\cite{maxwell1891theory}
While the two graphical languages are equally capable of deriving various partial derivative identities, the key difference is that sunray diagrams utilize vector sliding as its main technique, while Maxwell's diagrams utilize two-form sliding.
First, ``vectors are nicer than forms.''
Unlike one-forms, vectors have an advantage that graphical representation of their addition and decomposition is trivially easy.
Translating Maxwell's diagrams into mathematical expressions involves intricacies of differential forms (the ``reciprocal scaling'' behavior), but the vector language is free of such confusion.
Second, ``arrows are generally simpler than parallelograms.''
For example, compare the sunray-diagrammatic derivation of equation \eqref{eq:cp-cv} and that of Maxwell's, reproduced in Nash's article. \cite{nash1964carnot}
As a result, the sunray diagram is considerably less bulky than Maxwell's parallelogram sliding.

Also, the sunray diagram can provide complementary geometric intuitions and calculation pathways to the Jacobian technique.
The Jacobian technique\cite{crawford1949jacobian,carroll1965use,landau1980sp} can be thought of a more accessible alternative to Maxwell's area sliding, interpreting partial derivatives in terms of Jacobians.
It is perhaps the most competitive algebraic method to deal with partial derivative identities in thermodynamics.
However, for some occasions, such as equation \eqref{eq:chain2}, it is not easy to come up with the calculation pathway and apply the rules of Jacobians.\footnote{
Equation \eqref{eq:chain2} originates from the two-dimensional character of the symplectic form, $\delta^{d}_{a}\omega_{bc}+\delta^{d}_{b}\omega_{ca}+\delta^{d}_{c}\omega_{ab}=0$.
However, transforming this identity written in terms of Jacobians into the form of practical use (equation \eqref{eq:chain2}) requires several steps.}
In contrast, the sunray diagram derivation is simple and straightforward.
This is because the geometrical nature of equation \eqref{eq:chain2} is vectorial rather than bivectorial.
Some identities are most naturally understood in terms of oriented lines, while some others are most naturally understood in terms of oriented parallelograms.
This, at the same time, implies that a little practical value is gained from employing the sunray diagram for some cases.
For instance, regarding equation \eqref{eq:chain3}, some may find that the algebraic derivation (equation \eqref{eq:chain3_re}) is more straightforward to utilize than the diagrammatic derivation (figure \ref{fig:proof3}).
Nevertheless, both the Jacobian and the sunray diagram methods provide their own insights, being the canonical languages of two basic entities on the (two-dimensional) thermodynamic state space---oriented lines and parallelograms.

The visual intuition and aesthetic brevity of the sunray diagram make it appealing to utilize it in undergraduate education.
It does provide several visual justifications and explanations, perhaps the most impressive example being the peculiar minus sign in equation \eqref{eq:triple}.
The visualization of thermodynamic degrees of freedom as the dimension of the sunray diagram may help beginners to understand the difference between the number of variables and the actual thermodynamic degrees of freedom.
The sunray diagram effectively clears up beginners' misconceptions about the partial derivatives, providing a transparent and unambiguous picture.
Furthermore, users of the sunray diagram can easily classify and generate partial derivative identities, grasp the blueprint of their proofs, enhance their understandings on partial derivatives and differentials, and, even more, motivate themselves to enjoy exercising the ``sunray gymnastics.''

Lastly, as one of the future directions, the sunray diagram can be applied outside of thermodynamics.
In general, one can consider general systems of differentials with a symplectic structure or not.
Also, one can readily observe how the equations in Hamiltonian mechanics\cite{peterson1979analogy} translate to sunray diagrams.
Besides, since the current work concentrated on thermodynamic identities related to the basic variables, calculus of differentials involving further thermodynamic potentials interrelated by Legendre transformations remain to be explored in terms of the sunray diagrams.
We believe that this work not only introduces the new educational tool but also serves as a platform to explore various graphical languages of thermodynamics, promote understandings about the geometrical structure of thermodynamics, and enrich the graphical pedagogy in physics education.

\appendix
\setcounter{equation}{0}
\renewcommand{\theequation}{\Alph{section}\arabic{equation}}
\renewcommand{\thesection}{Appendix}
\section{\label{sec:App}Visualization Scheme for Tensors}
We provide a succinct introduction to the notation and the visualization scheme of multivectors and multiforms that are used throughout this article.
Our visualization scheme coincides with that of Schutz\cite{schutz1980geometrical} and Misner, Thorne, and Wheeler\cite{mtw2000gravitation}.
For the mathematical concepts themselves rather than the visualization rules, refer to sections 5.1--5.4 of Nakahara \cite{nakahara2003geometry} and pp. 113--167 of Schutz, \cite{schutz1980geometrical} as mentioned earlier in the opening part of section \ref{sec:Exp}.

A one-form $\lvec{w}$ is a linear map that sends a vector $\vec{v}$ into a scalar $\left<\lvec{w},\vec{v}\right>$. 
For this reason, a typical visualization of a one-form $\lvec{w}$ is an equally spaced parallel surfaces (with direction) where the number a vector $\vec{v}$ pierces them equals to the contraction $\left<\lvec{w},\vec{v}\right>$. 
We would like to write $\left<\lvec{w},\vec{v}\right>$  as $\lvec{w}\vec{v}$ in short.
Compare $\lvec{w}\vec{v}$ to $\braket{\beta}{\alpha}$, the contraction of a bra $\bra{\beta}$ and a ket $\ket{\alpha}$.
As bras are ``dual kets'' where the dual map being the dagger operation ($\dagger$), one-forms are also called ``dual vectors.''
Our notation that denotes one-forms with left-sided arrows is to hint such duality.

A two-form $\bm{\gamma}$ is a map that returns a scalar when a bivector is given as an input:
$\bm{\gamma}\,:\,\vec{u}\wedge\vec{v}\mapsto\left<\bm{\gamma},\vec{u}\wedge\vec{v}\right>$.
Therefore, it can be visualized as an ``egg-crate,'' a slanted lattice (with an orientation) so that the number of cells that a bivector covers equals the contraction of it and the bivector.\footnote{A bivector is an oriented area.
The simplest way to create a bivector is wedge producting two vectors---$\vec{u}\wedge\vec{v}$ is the oriented parallelogram that $\vec{u}$ and $\vec{v}$ generates.}
Two egg-crates are regarded as equivalent (i.e., representing the same two-form) if they have the same oriented area of a unit cell.
Lastly, it should be clarified that our convention normalizes the contraction between bivectors and two-forms as
\begin{equation}
    \left<\lvec{p}\wedge\hspace{-0.1em}\lvec{q} , \hspace{0.1em}\vec{u}\wedge\vec{v} \right>
    :=
    \lvec{p}\vec{u}\hspace{0.25em}\lvec{q}\vec{v}
    -
    \lvec{p}\vec{v}\hspace{0.25em}\lvec{q}\vec{u}
\end{equation}
for its ``area overlap'' interpretation to be valid. Meanwhile, in the spirit of our ``arrowheads as abstract indices'' notation (or syntax highlighting morphisms in the vector space category \cite{coecke2011new} in accordance with the Penrose graphical notation \cite{penrose1986spinors}), a $(0,2)$-tensor should be written with two left-sided arrows: $\lvec{\lvec{{\gamma}}}$.
However, we decided to write it simply as a boldface letter $\bm{\gamma}$ to avoid clutter.

\begin{figure}[t]
    \centering
    \includegraphics{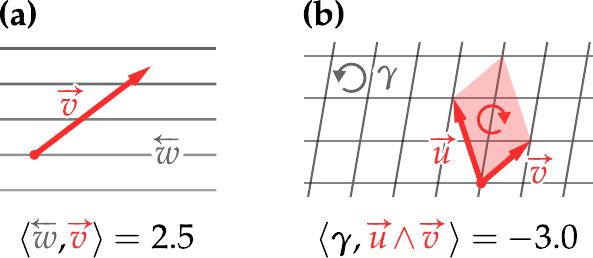}
    \caption{(a) Contraction between a one-form $\lvec{w}$ and a vector $\vec{v}$ equals to the number of piercings. (b) Contraction between a two-form $\bm{\gamma}$ and a bivector $\vec{u}\wedge\vec{v}$ equals to the number of $\bm{\gamma}$'s cells that $\vec{u}\wedge\vec{v}$ covers.}
    \label{fig:app0}
\end{figure}

Next, we discuss exterior calculus on two-dimensional manifolds.
To proceed in a low-cost way, one can assume an arbitrary coordinate system $(x,y)$ and its coordinate basis vectors $\vec{\mathrm{e}}_x$ and $\vec{\mathrm{e}}_y$, then later confirm the covariance of the equations.
The exterior derivative is denoted by $\d$.
It yields a one-form field $\d A(x,y)$ when acted on a scalar field $A(x,y)$:
\begin{equation}
    \d A = \pfrac{A}{x}{y}\lvec{\mathrm{e}}^x +\pfrac{A}{y}{x} \lvec{\mathrm{e}}^y,
    \label{eq:def_extd}
\end{equation}
while the functional dependence ``$(x,y)$'' is omitted for brevity.
$\lvec{\mathrm{e}}^x = \d x$ and $\lvec{\mathrm{e}}^y = \d y$ are called the coordinate basis one-forms and satisfy
\begin{equation}
    \lvec{\mathrm{e}}^x\vec{\mathrm{e}}_x = 1,\,\,
    \lvec{\mathrm{e}}^x\vec{\mathrm{e}}_y = 0,\,\,
    \lvec{\mathrm{e}}^y\vec{\mathrm{e}}_x = 0,\,\,
    \lvec{\mathrm{e}}^y\vec{\mathrm{e}}_y = 1;
\end{equation}
i.e., they form the dual basis of the coordinate basis vectors.
Note that $\d A$, defined by equation \eqref{eq:def_extd}, is a coordinate-independent object.
A typical visualization of the one-form field $\d A$ is contour lines of $A$.
This is because $\left<\d A, \vec{v}\right>$ equals to the first-order amount of change of $A$ for a displacement $\vec{v}$:
\begin{align}
    \left<\d A, \vec{v}\right>
    = \left<\pfrac{A}{x}{y}\lvec{\mathrm{e}}^x +\pfrac{A}{y}{x} \lvec{\mathrm{e}}^y,\, v^x \vec{\mathrm{e}}_x + v^y \vec{\mathrm{e}}_y\right>
    = \pfrac{A}{x}{y} v^x + \pfrac{A}{y}{x} v^y.
\end{align}

While equation \eqref{eq:def_extd} does not deviate much from the familiar calculus of differentials, the real game starts when differential multiforms are considered.
When a one-form field $A\hspace{0.07em}\d B$ is acted by $\d$, a two-form field is produced:
\begin{equation}
    \d \left( A\,\d B \right) = \d A \wedge \d B.
\end{equation}
Note that applying $\d$ to a constant scalar field gives $0$; hence,
\begin{equation}
    \d\left(\d A\right) = \d 1 \wedge \d A =0.
\end{equation}
The two-form field $\d A\wedge \d B$ is visualized as a curved array of parallelepipedal cells (a curved egg-crate) made by intersecting contour lines of $A$ and $B$.

In mathematics-oriented texts, \cite{mathoriented1, mathoriented2, docarmo2016differential, nakahara2003geometry} vectors and one-forms are not syntax highlighted, i.e., they appear without any particular symbols.
However, for physics students those who are beginners, it could be pedagogical to designate specific notations to vectors and forms.
Our notations $f$, $\vec{v}$, $\lvec{w}$, $\bm{\gamma}$, $\d$ correspond to Schutz's $f$, $\overline{v}$, $\tilde{w}$, $\tilde{\gamma}$, $\tilde{d}$, \cite{schutz1980geometrical} and Misner, Thorne, Wheeler's $f$, $\bm{\mathrm{v}}$, $\bm{\mathrm{w}}$, $\bm{\gamma}$, $\bm{\mathrm{d}}$. \cite{mtw2000gravitation}
For the symplectic gradient vector field corresponding to the one-form field $\d A$, we used the notation $\longvec{dA}$.
This is motivated by Schutz's $\overline{dA}$ (p. 170 of \cite{schutz1980geometrical}), while a more common notation is $X_{A}$.\cite{symplectic1, symplectic2, symplectic3, nakahara2003geometry}
It is important in our discussions to distinguish between ordinary differentials (infinitesimal amount of changes) and differential forms and see the correspondence between differential one-forms and symplectic gradient vector fields.
Our notation suits well to these purposes.
The right- and left-headed arrow notation is also the most efficient and intuitive way to present our idea of ``various implementations of an abstract syntax by supplementing it with concrete semantics,'' discussed at the end of section 3.3.

\end{document}